# Physical space and cosmology. I: Model

**Valeriy P. Polulyakh**

**Abstract** The nature of the physical space seems the most important subject in physics. A present paper proceeds from the assumption of physical reality of space contrary to the standard view of the space as a purely relational nonexistence - void. The space and its evolution are the primary sources of phenomena in Mega- and micro-worlds. Thus cosmology and particle physics have the same active agent - physical space.

## 1 Introduction

A Big Bang cosmology originally has been based on the following: GR, two hypotheses and two key observations. The first hypothesis was a Lemaître's "primeval atom". "At the origin, all mass of the universe would exist in the form of a unique atom; the radius of the universe, although not strictly zero, being relatively very small. The whole universe would be produced by the disintegration of this primeval atom. It can be shown that the radius of space must increase." [1, p. 86] "…"evolution", starting from an initial atom, may be explained by occurrence of an unstable equilibrium between two forces; the gravitation attraction and a repulsive force whose existence is suggested by the theory of relativity in connection with the cosmological constant." [1, p. 19] The second hypothesis was a Gamow's idea that the early universe was not only very dense but also very hot. An afterglow is the relict of radiation era of the early universe. "The Big Squeeze which took place in the early history of our universe was the result of a collapse which took place at a still earlier era, and that the present expansion is simply an "elastic" rebound which started as soon as the maximum permissible squeezing density was reached." [2] An observational Hubble law became a confirmation of the Lemaître's theory. The observation by Penzias and Wilson of the cosmic microwave background radiation (CMBR) was accepted as a confirmation of the Gamow's hot early universe.

Later the original Big Bang model acquired numerous innovations which made it not as transparent in a sense what is an essential and what is incidental. Here we do not intend to do any comprehensive analysis of these novelties. Instead, we address to the fundamental problems which, in our opinion, still are not solved.

*Singularity*. The Big Bang model does not provide the physics of the very beginning. From the passages cited above it is clear that neither Lemaître's nor Gamow's hypothesis have singularity in the beginning. The paradox is that singularity follows from the GR, however GR itself is invalid under the conditions it predicts. Since singularity is a failure or uncertainty of the cosmological model we need to find nonsingular conditions for the beginning.

*Second law of thermodynamics*. From the thermodynamical point a Lemaître's model with its cold and highly structured "primeval nucleus" that had low entropy is a more favorable than Gamow's universe in the state of thermodynamic equilibrium with the maximum thermal entropy. As long as Gamow's universe also has a maximum permissible density it has a maximum gravitational entropy. Thermodynamic equilibrium requires highly collisional condition which contradicts the low gravitational entropy. Therefore the full entropy was extremely high. It was a quite strangely designed universe that started in the most unfavorable condition. Logically, the universe with the minimum initial entropy seems more reasonable. It is not easy to understand the idea of the nature: why has it spent an enormous energy to create particles and antiparticles and then annihilated them to produce the maximum entropy in the form of CMBR photons? Penrose has argued [3] that thermalization increases the entropy, so invoking inflation is worse than useless for solving the 2nd law problem. The attempts to solve this problem by virtue of the expanding space do not look promising. It is like putting your backyard in order by throwing rubbish on the neighbors' lot. First, the physics of the space expansion, as we will see later, is a quite controversial subject. Second, to reduce gravitational entropy by scattering matter from highly collisional regime we need to perform the work that only aggravates the problem. Essentially, the problem is: if CMBR is a product of the beginning where has the entropy accumulated since then? The most fashionable answer is: this entropy is hidden in the black holes. However, this is analogous to explaining an existence of the black cat in the dark room by the darkness in the room. The existence of the black holes themselves is still the hypothesis, though very popular one.

*The law of conservation of energy*. Both Lemaître's and Gamow's models nominally maintain this law. However already in the Gamow universe we encounter problems. For instance, the CMBR photons lose energy due to redshift effect. Where does this energy go? Let us appeal to the cosmology textbooks. "… the net radiation energy in closed universe decreases as 1 / a (t) as the universe expands. Where does the lost energy go? … there is not a general global energy conservation law in general relativity theory." [4, p. 139] The expanding universe does not perform work so why is its temperature drops? "The conservation-of-energy principle serves us well in all sciences except cosmology. The total energy decreases in an expanding universe. Where does energy go in an expanding universe? The answer is nowhere, because in the cosmos, energy is not conserved." [5, p. 349] Sometimes zero-energy principle is used: a continual adjusting of the matter and gravitational contributions to make the sum zero. In regards to the reddening of the CMBR photon we need specific physical mechanism that realizes this adjustment of the energy between photon and gravity. What mechanism may act upon it? If it is a gravitational frequency shift

$$d\nu / \nu = -d\Phi / c^2$$

where Φ is a gravitational field potential then a result depends of potential difference. Since a real matter distribution has non-uniformities up to hundreds Megaparsecs the frequency shift has to have random component. What is more important the average shift on cosmological scales probably has to negate systematic reddening. Both effects are not observed. The field of view of the relevant literature shows that: 1) there is no general principle of energy-momentum conservation in GR; 2) the stress-energy of the gravitational field is well-defined neither locally nor globally.

*Gravitation theories*. Despite development of enormous number of relativistic and non-relativistic gravitation theories there is no one established theory that embraces all relevant phenomena. It is impossible, even superficial, to attempt to fully examine this boundless subject. However, let us mention here a little known issue concerning the role of pressure in generating the gravitation field. Recently some theoretical objections were presented [6] on this role in the theories of gravity, which use the energy-stress tensor to represent matter. Does the pressure gravitate? To my best knowledge nobody tested it experimentally yet. This is extremely substantial issue since the gravitation properties of the negative pressure is an important component of many fashionable cosmological models. There is some hope on the quantum gravity; however there are no any experimental phenomena about gravity's quantum nature. Also, if gravitons exist we can detect them and distinguish between gravity and acceleration that contradicts equivalence principle. Most likely the nature of physical space has to be object of theoretical study. Then we can hope to find the nature of gravity as a consequence of this study.

*Expansion of the universe*. In spite of seeming simplicity of the phenomenon there is an arena of a stubborn discussion concerned with a redshift interpretation. The reader can easily find the relevant works. There are two different approaches: expansion is the motion of the bodies in space that is a kinematical expansion and expansion of the space itself. "Expansionists" are brief: "The expansion of the universe is a natural feature of GR" whatever it means. [7] In the language of Riemannian geometry expansion is an increase of the distance between points of space measured intrinsically. What is physically meant by "points" in empty space (GR is equally working in empty universe) is not clear however. Meanwhile "Anti-expansionists" are more inventive. "In an empty universe, what is expanding? Unlike the expansion of the cosmic substratum, the expansion of space is unobservable. Expansion of space on the physical level suggests the existence of a new mysterious force. If so, one can also expect non standard effect on small scales. Expansion of space is somewhat a matter of philosophy and semantics, rather than hard science." [8a] "There is no local effect on particle dynamics from the global expansion of the universe. Expanding space is in general a dangerously flawed way of thinking about an expanding universe." [8b] "The cosmological redshift is not an "expansion effect", in special cases it can be separated into kinematic part and a static (gravitational redshift) part. All our test equipment and comparisons are built of or use electromagnetic forces, and they should also expand with the universe; so any cosmological redshift would be undetectable in principle. If one looks at space itself apart from the associated fluid of cosmological matters, it is not at all certain what the expansion of space means." [8c] It is very important for us to explain the increase of the global volume of the universe without stretching of the local space. We have to comprehend a physics of this controversial phenomenon.

*A formation of the cosmic structures*. By cosmic structures we mean not only a large scale network but all the objects populating the cosmos: the stars, star associations and clusters, gas bubbles and shells, galaxies and quasars, star- and galactic- jets, groups and clusters of galaxies, and finally the large scale structures of the cosmos. Do we have a unique cosmological picture of creation and evolution of this diverse zoo? This question needs elucidation. The answer depends on what we understand under cosmology. Where is the boundary between cosmology and the rest of astrophysics? For example, are the protogalactic gas clouds in the province of the cosmology or astrophysics? From the point of cosmology these clouds obviously exist, however an observational astrophysics has never seen them. In some sense the difference is as follows: what we really see is astrophysics, however what we believe we could see is a cosmology. In reality all this zoo is a territory of a dozen of astrophysical disciplines. Apparently all the knowledge collected to date has to be strung on some united idea. The question is who has to amalgamate this information? A modern cosmology, possessing itself as a science concerning with the largest scales and the earliest moments, does not cover all the spectrum of cosmic structures to present a unified frame. Another drawback of the current cosmological model is a demand of non detectable dark matter and dark energy.

*Infall – Outflow Problem*. There is a subject that astronomers are not disposed to discuss. The observations show availability of outflows and jets from the variety of cosmic objects starting with young stars and end with quasars. The nature of jets is not well defined: exploiting models from molecular and plasma streams to Poynting flux. Visually jets from parsecs to Megaparsecs scales look very similar, however the models of central engines from exotic magnetic configurations to black holes are used for their explanation. How can ubiquitous outflows support the hypothesis of accretional creation of stars and galaxies is not clear. Of course some weak accretion exists but its observational rate is absolutely insufficient for creation of the stars and galaxies.

*Observational evidence*. What is observational evidence in astrophysics? Let us recall an example of discovery of quasars. Maarten Schmidt defined them as the star-like objects with high redshift. Even if this definition was not absolutely strict it is extremely operational and allowed to find thousands of new cosmic objects. As an opposite example let us take the CMBR. There is an obvious and pronounced dipole component that has very clear physical nature. However the remaining ripples that are measured with COBE, WMAP and now with Planck are very dependent of particular instrument, its calibration, artifacts of signal processing, removing galactic foreground contamination and so on. There is nothing wrong in the process of measuring itself. What is unprecedented is a reckless use of this controversial statistical material as a main observational foundation of the picture of the World. It may happen that structures now visible with Planck and being treated as foreground are indeed of first importance. Another example is observation of supernovae redshifts which changed astronomer's believes overnight from decelerated universe to accelerated one. May be it is right, may be it is not. However statistical data require very cautious interpretation and taking into account availability of contrary results [9]. In the work [10] the analysis of center of the Galaxy near Sgr $A^*$ showed that most of gas is found to outflow from the region. How can we conclude that there is a black hole? One more example. If we believe that the universe started from the uniform cloud of neutral hydrogen then we have to see stratification continuous in time of the gas clouds mixed up with the young galaxies. This picture has to show a gradual parallel increase of the galaxies concentration and diminution of the gas clouds. Astrophysicists have never yet seen this metamorphosis. How do we know that our picture of cosmic evolution is correct? Also the structures of large galaxies show little relation to their environment, contrary to the standard picture of assembly of galaxies.

So, if we believe that our World has started sometimes ago we are still in the position to decide which hypothesis, Lemaître's or Gamow's was closer to reality. There is an opinion that the problems in the standard cosmology could be solved by adjusting of details. Our suggestion is that we have to go back to the conceptions and use the observations accumulated since. In the works [11] I have tried to construct a new paradigm based on the idea of continuous physical evolution of space. The rest of this paper is a short sketch of this idea; its consequences will be presented in the next paper.

## 2 Physical space

As a rule, the cosmological models are constructed on the basis of the theory of gravity. It makes sense since the behavior of space-occupying matter is of the greatest interest. But where do this matter and space come from? It seems there is a missing link between cosmic structure and gravity. The structure and evolution of space are the central problems of physical cosmology. What is a physical space? We need a short historical excursus in philosophy and metaphysics first.

The most famous debates about philosophical controversies of the space are presented in the exchange of papers between Leibniz and Clarke, a friend and disciple of Newton. This is so called "The Leibniz-Clarke correspondence" [12], where Leibniz produced arguments against Newton's absolute space and Clarke argued for Newton's conception. Leibniz argued that space can exist without the mind but it can not exist in the absence of matter: "I hold space to be something merely relative … Motion and position are real and detectable only in relation to other objects … therefore empty space, a void, and so space itself is an unnecessary hypothesis." Newton's position (in the modern language) was the following. I. Absolute space is objectively real, distinct from material bodies and exist without the mind; II. Space is a similar and immovable pseudo substance ontologically independent of things and existing without relations to bodies ("Although space may be empty of body, nevertheless it is not itself a void and something is there because spaces are there, though nothing more than that." III. Space can exist in the absence of matter, but matter cannot exist without space. IV. The entities of absolute space are real points and places while material bodies occupy these places and points. V. Space is continuous, infinitely extended, has internal metrical structure and possessed Euclidean geometry. The Newton's line of thinking about absolute space has been transformed into "substantivalism" (space is a uniform entity, container, existing objectively and independently of its material content, though Newton himself never insisted that space is a substance) and a Leibniz's turn of mind became "relationism". William Clifford (1845-1879), in these terms, was adherent of "supersubstantivalism" since he supposed space as a first-order substance: "That this variation of the curvature of space is what really happens in that phenomenon which we call the *motion of matter*, whether ponderable or ethereal." [13]

Newton advanced two famous arguments for absolute acceleration: a bucket and the rotating globes though experiments that allow us to consider an existence of a topological absolute space. Another topological consideration for absolute space is a famous Kant's enantiomorphism of the hands.

Introduction of space-time in SR and GR does not bring something new in this controversy since S-T should be supposed as a real entity like absolute Newtonian space. H. Alexander observed [12]: "… the GR does seem to imply the reality of space-time and meaningfulness of speaking about its properties … To some writers it has seemed that when in the Correspondence Leibniz criticizes the concept of absolute space and time, he anticipating Einstein. On the other hand, Leibniz's fundamental postulate is that space and time are unreal. No one therefore would have rejected more strongly than he a theory which ascribes properties to space-time."

Einstein did not established in his theories a relationism, as many think, quite an opposite: he needed something real for space because the rest in his constructions is the tool. Indeed, what are the SR and GR? These are imaginary rods and clocks plus Minkowski pseudo-metric and a tensor calculus developed by Ricci. In GR Einstein for the matter had the energy momentum tensor but for the space he had only a mathematical construction four-dimensional manifold (M) and Lorentz metric (g) that does not meet requirements of a metric but it is a pseudo-metric. To be a physical theory, this mathematical conglomerate needed some portion of reality. This is exactly why Einstein persistently tried to construct reasonable foundation for the “new ether” [14], though his goal was apparently comprehension of physical space.

It is doubtful thus whether debates between substantivalism and relationism may bring a progress in our comprehension of the nature of space, so we have to appeal to the nature itself. There are two distinct notions of space [11]:

1. The referential space that is a part of space-time structure and depict the world of events.
2. The physical space that is a background of the world of events and endowed with the measurable physical properties.

What are these properties? Here are some of the measurable properties.

**Euclidean Metric**. The space we live in is a flat one with a clear Euclidean geometry that is a base for the conservation laws.

**Physical space is homogeneous and isotropic**. It is precisely these properties that explain why physical space itself can not be a referential space.

**Expansion of space**. It is hardly to believe in expansion of the space if it does not exist. The expansion rate or Hubble constant $H_0 = 72$ km $s^{-1}$ $Mpc^{-1}$.

**Gravitational properties of space**. The ability of space to transmit attraction from one mass to another is an obvious property of space that is measurable as Newtonian constant of gravitation $G = 6.674 \times 10^{-11}$ $m^3$ $kg^{-1}$ $s^{-2}$.

**Electromagnetic properties**.

The electric permittivity of free space $\varepsilon_0 = 8.854 \times 10^{-12}$ F $m^{-1}$
The magnetic permeability of free space $\mu_0 = 1.257 \times 10^{-6}$ N $A^{-2}$
Characteristic impedance of vacuum $Z_0 = 376.730\ \Omega$
The speed of light in vacuum $c_0 = 299\ 792\ 458$ m $s^{-1}$.

**Lorentz invariance**. This property is most controversial. There are two different approaches to SR: “principle” theory of Einstein and “constructive” theory of Lorentz and FitzGerald. The Lorentz invariance is a property that is common to both approaches. This property was central to the Poincare’s relativistic model. Generally there are two possibilities of relations between inertial frames: Galilean with $c = \infty$ and Lorentzian with the fixed finite c. It is clear that the nature picked out a second possibility even though the first one is much simpler. Why? A simple answer is because a relation $E = mc^2$: infinite speed requires infinite energy in the presence of the mass. Thus the nature found an elegant solution. She kept masses but based c value high enough to keep our habitual World simple, practically Galilean with lucid Newtonian physics. The relativistic phenomena only become apparent near the speed of light that is very seldom in our daily experience. Therefore the nature made our lives much easier than it could have been. The price of this concordance is imperceptibility of the physical space structure by simple experiments. Physical space does not represent an absolute frame for electromagnetic phenomena.

Let us consider the distance between emitter (E) and receiver (R) of the light is 1 million km. The speed of light depends of the properties of the medium between E and R as $c = (\varepsilon\ \mu)^{-1/2}$. The time of flight of photon t is different depending of the medium. For vacuum t = 3.33 sec, for glass t = 5 sec, for diamond t = 8 sec. In the vacuum we can not consider $\varepsilon_0 \rightarrow 0$, or $\mu_0 \rightarrow 0$ for then $c \rightarrow \infty$. So from the Maxwell theory c depends of the properties of the free space or physical vacuum. However, Michelson–Morley experiment did not reveal any structure of the vacuum. Does it mean that physical vacuum is not a medium? It means that this medium possess the Lorentz invariance property. The absence of visible interaction of photon moving relative to this medium may be roughly demonstrated by following experiment. Let us consider the motion of the hockey puck between two players. If the ice starts to move transversely to the puck's trajectory as a river we can not see this motion with the puck's traveling because negligible friction between the ice and the puck, though ice is a sufficient part of the physical system. So, the M-M experiment is too rough to probe the vacuum machinery; it is not appropriate for investigation of the delicate structure of physical space. Can we peep into the fine machinery of the physical space to see it's structure so carefully protected by the nature from our eyes? I think so. Free space, quantum mechanical vacuum, unified vacuum, "Dirac sea", quantum foam and topological defects with vacuum energy, virtual particles or vacuum fluctuations and decay of vacuum – all of these are entities and processes reminiscent of this stage that is the physical space. The vacuum polarization, observed experimentally [15], demonstrated that vacuum is not completely empty. The properties of quark–gluon systems also imply an existence of some vacuum structures: "… the most profound and surprising result to emerge from late twentieth-century physics may be the realization that what we perceive as empty space is in reality a highly structured and vibrant dynamical medium". [16]

It would be very strange if something that possesses physical qualities did not exist. However if something exists and own properties then it is *an entity*. The barrier for understanding this truth is a painful succession of the long standing battles between supporters and refuters of the luminiferous aether [aether = ether]. In his paper "Is There An Aether" P. Dirac claimed: "… with the new theory of electrodynamics we are rather forced to have an aether". We can also refer to P. Drude [17]: "The conception of ether absolutely at rest is the most simple and the most natural, -at least if the ether is conceived to be not a substance but merely space endowed with certain physical properties. "

Thus under physical space (PS) we comprehend the real medium endowed with certain measurable physical properties. I would like to call it milieu.

## 3 The problems

What are the problems that we have to address here? Some of them are already submitted in the Introduction. A numerous observations exclude the steady state model. Then we need the nonsingular conditions for the beginning. The very early world has to have as low entropy as possible with the CMB entropy being accumulated during the world's life. The flatness of space we live in has to be a natural state. What is the difference between Euclidean flat space and Einsteinean flat space? Euclidean one is simply naturally flat but for Einsteinean one to be flat we need unseen dark matter and dark energy. For Euclid and Newton the flatness was a natural condition of milieu. However in the current cosmological model flatness, which is obvious in observations, is a result of special hypotheses. That is too high a price. The Euclidean flat space and Minkowski space-time metric are sufficient to hold the conservation laws. We are keeping 3D space and time firmly as two separate concepts, connected by Lorentz invariance but physically essentially different. The model has to explain physics of controversial phenomenon - an increase of the global volume of the universe without stretching of the local space. The model of physical space must establish a connection between macro and micro worlds. In accordance with the Occam's razor this model should account for formation of the cosmic structures in the unified manner. Finally, as a bonus, the model ought to be capable to explain new, unexpected in the frame of standard cosmology (SC), observations. It is natural that the current state of our model can not account for the solution for the whole set of these requirements; it only outlines the plan of the necessary task that lies ahead.

## 4 The superfluid physical space.

If we agree that PS is a real medium then we have to study its properties. However, first we have to explain why we are talking about a space but not a space-time as is common in modern studies. It can be roughly explained as following. We can bite off piece of hamburger, this is analogous to "static" space-time treatment, but we can also bite off pieces of beef and bread separately and chew them, this is analogous to "dynamical" space and time treatment. And second, our interest is not a referential space but background one, so we treat PS as an entity but not the time.

Let us recall the treatment of expansion of our world in the SC. As we mentioned already there is a problem: the global increase of the world's volume doesn't have any effect locally. Let us imagine an elastic bag that embraces our world. The pressure inside will drop and the volume will increase. In the real world's case there is no local "pressure drop" but volume of the world increases. We can solve the problem in the following way. Let us assume that the bag is not elastic, it can expand without stress and it is filled with the fluid which pressure equal to the pressure outside the bag. How we can get expansion of the bag? One obvious way is to have inside containers with pressured fluid. If we open containers the outlawing fluid increases the volume of the bag but locally nothing changes.

We consider in our world the presence of two physically distinctive areas: 1) standard physical space with particles, gravity and other physical fields, 2) "compressed" space with non-zero "elastic" energy and lack of particles. At the expense of "elastic" energy the mass (dm) and physical space volume (dV) can be created. Our hypothesis is proportionality between dm/dt and dV/dt:

$$\Pi \frac{dm}{dt} = \frac{dV}{dt} \qquad (1)$$

where Π is a coefficient proportionality that have dimension [$m^3$/kg]. When the mass in the volume V is created then a new space outflow is proportional to the rate of density of matter $\rho$ creation

$$\Pi \oint_V \frac{d\rho}{dt} = \oiint_S \; (\mathbf{v} * \mathbf{dS})$$

where **v** is velocity of PS and by applying the divergence theorem we obtain

$$\Pi \frac{d\rho}{dt} = div\, \mathbf{v}. \qquad (2)$$

There is no dragging force applied to the stationary moving matter from the PS which possess properties of perfect fluid:

a) The PS has zero viscosity
b) There is no heat transfer in the PS flow
c) The motion of PS is always potential, so

$$curl\, \mathbf{v} = 0, \qquad \mathbf{v} = -\, grad\, \varphi \qquad (3)$$

where φ is the velocity potential.
The microscopic hydrodynamics of the superfluid physical space may permit only acoustic waves. Then there may exist "space phonons" witch are the carriers of gravity. However, this is very speculative reasoning and I do not see the way to prove it.

All the physical forces are functions of the relative coordinates not the absolute positions. Why? Our point is that inertia is caused by presence of the physical space. If the velocity of matter relative to PS is constant there is no force from the side of PS as it is in the case of motion of the d'Alembert's sphere relative to the perfect fluid (Fig.1)

The velocity field is symmetrical on the upstream and downstream sides of the sphere, consequently the total momentum of the flow is undisturbed. Since the flow cannot transfer any momentum to the sphere. So, the net force experienced by the fluid on the sphere is zero. As far as the velocities are symmetrical on the upstream and downstream sides, the pressures are symmetrical too, that follow from Bernoulli's formula $\rho v^2/2 + p_0$ = constant, where $p_0$ is a static pressure. It means that the sphere moves stationary through fluid without resistance.

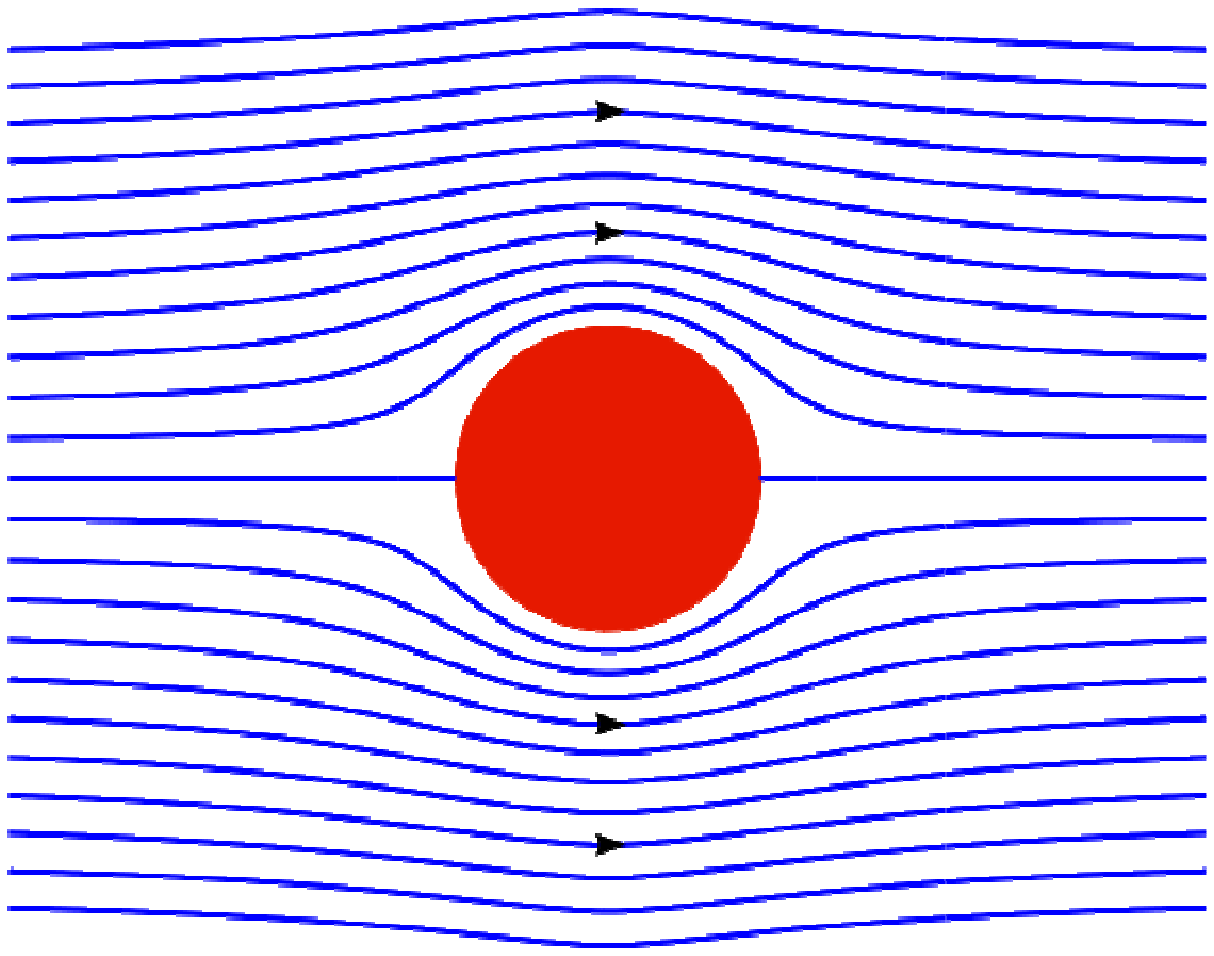

Fig.1 G.K. Batchelor (1967) "An introduction to fluid dynamics", Cambridge Univ. Press, page 424. (Wikipedia)

However, non-uniform motion (accelerated or slowed down) relative to the PS begets new force applied to the matter from the side of PS. The velocity field ceases to be symmetrical; so do pressure, therefore there is misbalanced force. This is the physical cause of inertia. Thus, unlike to the Mach, who believed that "distant bodies of our universe should contribute the most to the inertial forces", we dare say that inertia is a result of the local interaction of the matter with the PS. So, in uniform motion PS has no manifestation; only non-uniform motion relative to PS is displayed as inertia. If physical particle is moving uniformly there is no any interaction with the ambient PS but to accelerate particle relative to the PS we have to apply external force. Because of the vector nature of acceleration the appearance of inertia applies not only to a change of the magnitude but to the directional changes also. So, in the Newton's bucket experiment the inertia is caused by the interaction with the ambient PS. According to the 3rd Newton's law if PS itself moves non-uniformly relative to the matter it experience the force acting on the space from the matter side. In this case the PS gives rise to acceleration of the involved matter relative to the remote observer. In the case of the charged particles expended energy is released in the form of electromagnetic radiation that is excitation of the PS. We have to emphasize: *The uniform relative motion of the physical space and matter does not cause interaction because PS is the perfect fluid.* This is the answer on the question: "Why we do not feel the space?" However the recognition of the space by the presence of inertia was not clear. *The property of the particles to possess inertia is a manifestation of existence of the physical space.* A charged particle that moves uniformly doesn't spend energy because there is no disturbance of the PS and, as a result, there is no radiation.

An acceleration relative to the PS is also an acceleration relative to the bodies submerged into this space so it looks incorrectly that the inertia is caused by the acceleration relative to the bodies not to the PS. Let us imagine two small balls in the cosmic void. The other things like stars are on the millions of light years from the balls. Suddenly we apply some force to the first ball to accelerate it relative to the second ball. We can fill inertia of the ball. Who had created this inertia? Second ball is too small. The stars are too far away. The only agent is available for causation is the PS into which the ball is submerged, if we do not suggest long-range action.

All the forces are functions of the relative coordinates not the absolute positions. However an acceleration of the body needs the application of the force. Thus inertia is caused by presence of the physical space.

In Newtonian system of the World all the volume elements of the true space do not move relative to each other, the volume of this space is frozen once and for all. Here we will be guided by equation (1). It is obvious from this equation that the volume elements that do not contain matter production are not expanding. So, the space in our room is not expanding. Brooklyn is not expanding, New York is not expanding and the Earth is not expanding. The World contains enormous number of sources that causes uniform expansion. We can see expansion as a Hubble flow. Where does the uniform Hubble flow start? Since the Hubble expansion is a summary outcome of production of the flat space from the number of the sources the uniformity of expansion starts on the scales where density of (dm/dt) becomes homogeneous. It is a volume element comparable with the cluster of galaxies. But non-homogeneous expansion, which is not a Hubble flow yet, starts in the volume elements containing at least one star with an active sources. To understand why we do not see flying away of stars in the galaxies and of galaxies in the small groups we have to turn our attention to mechanics.

Proceeding from our comprehension of space the velocity of each body $\mathbf{v}$ consists of two components. One is a local peculiar velocity $\mathbf{v_{pec}}$ and other is a global carrying velocity $\mathbf{v_{car}}$ :

$$\mathbf{v} = \mathbf{v}_{pec} + \mathbf{v}_{car} \qquad (4)$$

Certainly **v** is a relative velocity whose radial component is depicted by the Doppler shift of moving emitters. The $\mathbf{v}_{car}$ is caused by increasing the volume of physical space and usually in observational astronomy is called by the recessional speed $\mathbf{v}_{rec}$. The local velocity $\mathbf{v}_{pec}$ is an equivalent of Newtonian "true" velocity. Since, as we mentioned already, Newtonian "true" space and our physical space, themselves, can not be the referential space we have to measure this velocity relative to its substitute. In this case the CMB is the substitution that outlines the physical space. So, by our definition $\mathbf{v}_{pec}$ is measured relative to the local CMB frame and $\mathbf{v}_{car} = \mathbf{v}_{rec.}$ Any velocity **v** in the World consists from the velocities of two types: 1) peculiar ≡ true ≡ local and 2) carrying ≡ receding ≡ global.

Since Newtonian true space is frozen it does not allow for an existence of the velocity of type 2. Newton stated:

> "Again, true motion suffers always some change from any force impressed upon the moving body…" [18]

And this is correct with respect to our peculiar or local motion. But, in contrast to Newtonian frozen true space, here is another type of motion which is a consequence of our model with the creation of the new volume elements.

Here is a significant difference between two pictures of motion. The carrying (or global) motion, unlike peculiar (or local), does not require the local force applied to the body to undergone a change. The physical space, just as Descartes "hydrodynamic" space, carries body without application of physical forces to this body. The physical space has no dragging effect relative to the bodies but it is a causal agent of inertia. Namely property of the inertia links particles and physical space. The inertia becomes apparent only when body accelerates relative to physical space. The essence of inertia is in the following. Since physical space cannot be a reference system there is a need for some agent which links particles with the physical space. An inertia is exactly this agent. In the Newtonian mechanics $\mathbf{v}_{car} \equiv 0$, therefore Newton's Second Law is always implemented for ($d\mathbf{v}_{pec}/dt$). An emergence of movable physical space alters Newton's Second Law. Now with $\mathbf{v}_{pec} = \text{const}$ particles can undergo relative acceleration

$$d\mathbf{v} / dt = d\mathbf{v}_{car} / dt \qquad (5)$$

at the expense of global motion. Let us consider 3 galaxies that stand out from each other at distances: $\mathbf{r}_{12}$, $\mathbf{r}_{23}$ and $\mathbf{r}_{13}$ which are connected by relation:

$$\mathbf{r}_{12} + \mathbf{r}_{23} = \mathbf{r}_{13}. \qquad (6)$$

The relative velocities of these galaxies: $\mathbf{v}_{12}$, $\mathbf{v}_{23}$ and $\mathbf{v}_{13}$ are connected by relation:

$$\mathbf{v}_{12} + \mathbf{v}_{23} = \mathbf{v}_{13}. \qquad (7)$$

In general, these relations have no correlation for the galaxies chosen at random. Now let us consider more and more separated galaxies. Then, in average, $\mathbf{v}_{pec}/\mathbf{v}_{rec} \rightarrow 0$ and hence

$$\mathbf{v}_{rec,12} + \mathbf{v}_{rec,23} = \mathbf{v}_{rec,13}. \qquad (8)$$

For high separation between galaxies an increase of volume between them is proportional to distance and isotropic. So, recessional velocities connected with distances as follows

$$\mathbf{v}_{rec} = H\mathbf{r} \qquad (9)$$

that is a Hubble law. Therefore Hubble law starts at the distances where the density of (dm/dt) becomes homogeneous. The peculiar velocities were averaged out by virtue of their randomness and smallness. In the process of successive cascading down of the compressed space there is a moment when the production of physical space becomes sufficiently uniform. Beginning from this time a Hubble law implemented for the scales above the cell of non-uniformity. The matter on these scales freely follows to the recession of space but on the scales less then the size of the cell the matter is susceptible to the physical forces, basically gravity. Since the uniform motion of the physical space does not hold dragging effect or viscosity, these forces can create acceleration relative to physical space and bound the particles near the gravitational masses.

If none of the forces act on the particles they retain their motion with the volume element of physical space that is an inertial effect. If the volume element changes its motion, for instance, it accelerates under the influence of surrounding volume elements; the particles retain their motion relative to the volume element that can be detected relative to CMB which, in its turn, retain motion relative to the volume element. The driving force of the changes of motion of the volume elements is an action of elastons by production of new physical space. In this sense the inertia is a manifestation of the existence of real physical space. An acceleration of the physical space caused an acceleration of the particles relative to the remote observer.

It is obvious that introduction of global static referential system is no possible in consequence of the creation of new volume elements of the physical space. Only local referential systems can exist in the volume elements without compressed space. These parts of the physical space are the reminiscence of the Newtonian true space where Newtonian and SRT physics is applicable.

There are two instruments for the measuring of the motion. The Doppler Effect is suitable for the radial relative motion $\mathbf{v}$ of the particles that emit the electromagnetic energy and the CMB dipole for the local velocities $\mathbf{v}_{\mathbf{pec}}$.

## 5 The elastons and evolution of PS

The central idea of a new paradigm is an existence of a certain entity – the elaston, which can give rise to the physical Euclidean space, matter and energy all the way along its evolution. There are no spatial singularities.

There is the question in any cosmological model: “Why, after an infinity of nonexistence, was the World generated, at one moment rather than another?” In this model the answer is simple; the existence of our World is only a moment in the perpetual evolution of space. The elaston emerges as a link in the infinite transmutation of space so it is not a peculiar *moment* in cosmic time. Two conceptions lie in the foundation of elastonic hypothesis:

1. An existence of specific elasticity as a property of physical space.
2. A concentration of the finite elastic energy of space in the form of soliton-like entities – elastons which are a domains of "compressed" space.

Under soliton we comprehend here a long living spatial formation protecting its shape against spreading. Clifford's idea of representation of matter as a space curvature was rejected from the very outset. The matter is considered as a by-product of evolution of space however the energy of elastons is described in the geometrical manner trough metrical properties of space. So elaston is not the matter, in the sense of particles, but it is pre-matter that can give birth to particles. Though our surrounding or milieu we call "physical space" the space inside the elaston is also real but we call it "elastonic space". More detailed description of elastonic space was given in [11] therefore here we give only its main properties. We will act in the following way – using our understanding of reality try to find promising mathematical resources and as Dirac recommended "…try to interpret the new mathematical features in terms of physical entities…" At the same time we must endeavor not move away from the physical reality. We agree, this way is controversial thought may be fruitful under strong observational control. In regard to observations, we will concentrate on them in the next paper, when we compare corollaries with astrophysical phenomena. New mathematical tool we have chosen already for our model is a Ricci flow technique. We have to start with non-metrizable space.

The fact is that we have no intuition about non-metrizable spaces. We don't even have physical language to discuss such a reality. Only thing we can do is to use mathematical terminology relying on the hope that in the future we could develop suitable physical comprehension.

In the beginning of the chain we can put so-called pathological space that is non-metrizable. A domain of well-behaved space arises as a result of self-perfection. Well-behaving is a continuous process, one can have space with better and better behavior. On this stage the space doesn't have any geometry. It is analogous to antichain where any two elements are incomparable. Well-behaved domain evolves into space that is second countable. If for all this the points of the space can be separated by neighborhood then evolution has led to metrizability of space. Our chain may now have a form:

Pathological space → Domain of well-behaved space → Second-countable space → General Riemannian space → Ricci flow → Elaston → Particles + Euclidean space

It is unclear whether this chain extends in one direction, in both directions or is that all. It is appropriate to recall here a metaphysical question of Aristotle about our World:

> "This is held in the Timaeus, where Plato says that the heaven, though it was generated, will none the less exist to eternity. … why, after always existing, was the thing destroyed, why, after infinity of not being, was it generated, at one moment rather than another?" [19]

We have a simple answer in our model. Aristotle also suggested an interesting idea:

> "From our arguments then it is evident not only that there is not, but also that there could never come to be, any bodily mass whatever outside the circumference. The world as a whole, therefore, includes all its appropriate matter, which is, as we saw, natural perceptible body. So that neither are there now, nor have there ever be formed more heavens than one, but this heaven of ours is one and unique and complete. It is therefore evident that there is also no place or void or time outside the heaven. …Hence whatever is there, is of such a nature as not to occupy any place, nor does time age it; nor is there any change in any of the things which lie beyond the outermost motion; they continue through their entire duration unalterable and unmodified, …" [19]

A conception of Aristotle virtually signifies that the space, out of our World or the domain of space where we live, is non-metrizable.

The understanding of properties of physical 3-space is our main concern. It is not possible to equip the empty homogeneous physical space with concept of distance and hence to ascribe any metric to this space. An introduction of one particle retains uncertainty which may be removed by introduction of at least two particles. In the elaston an uncertainty is removed by non-uniformity with characteristic distance scale $s_0$. So, the logical chain of this model is as follows: “non-uniformity → distance → metric → elaston with elastic energy → Euclidean space with particles”. The created Euclidean space *certainly* contains particles which removed uncertainty in scale and induced a reference system.

Even if our primary interest is our World, we have to consider space lager than World. But this leads to the boundary problem. Plato and Aristotle believed that our World is finite and has boundary. A modern cosmology is zealously ignoring this possibility assuming only spaces without boundary like Riemannian “spherical” space or 3-sphere. This is notwithstanding the lack of evidence and proceeds solely from philosophical conception “Copernican Principle”. An idea of infinity bears no relation to reality but it is only arbitrary extrapolation of the milieu.

Let us consider the following thought experiment. Assume the flat 3-space that is undergoing a non-uniform squeezing. For example, squeezing creates a bell-shaped distribution of a variable scale factor a (t, r) where t is a time and r is a space coordinate. Let us suppose Euclidean 3-ball and mark small balls with equal radii (dr) along the radius of a big ball. Now let us squeeze a big ball to the center. The squeezing does not depend of the angular coordinates but does depend on a distance from the center. So if “s” is a distance from the center of the ball, then all physical sizes are proportional to some quantity “a” that is a scale factor. Scale factor is a coordinate dependent function. Each of the originally equal small balls now has different volume. It means that this space has Max squeezing in a center and a scale factor “a” is a ratio $a(s) = dr(s) / dr(s{=}0)$, where $dr(s{=}0)$ is a radius of a small squeezed ball in the center and $dr(s)$ is a radius of a small ball on the distance “s” from center. So, all geometrical sizes are increasing as $a(s) = \cosh(s/s_0)$ from center, where $s_0$ is a characteristic size of a big squeezed ball. It is squeezed in a center more than on the periphery. It is necessary to conceive that a given procedure of compression of space is a conformal (or quasi-conformal) transformation. This compression preserves both the angles and the shapes of infinitesimally small domains of space and only changes their size. The infinitesimally small balls transform in the balls of different radii (or the infinitesimally small balls transform into infinitesimally small ellipsoids in which the ratio of major axis to minor axis is bounded in the quasi-conformal case). A scale factor depends on the location, but not on direction. Our compressed metric 3-space is conformally flat that is a Cotton tensor is vanishing.

Now we use a ruler to measure the length of geodesics of this entity on some level and the diameter on the same level. Our measurement show that the diameter is bigger than necessary for a flat space. We have “positive curvature” by definition. Or, another way, if we calculate volume of this ball V then it is less than volume $V_0$ of the ball with the same radius $r_0$ in Euclidean space. According to Bishop’s inequality $V < V_0$ this entity has a positive Ricci curvature that describes curvature of 3-space completely. Ricci curvature is an intrinsic curvature that does not require embedding in the surrounding 4-space. Because Ricci curvature controls the growth rate of the volume of metric balls the more squeezed entity that has higher curvature may experience higher expansion of the volume when it approaches Euclidean space. We call this entity – the elaston.

First elaston was a bounded space (3-ball) surrounded by non-metrizable topological space. It means that the complementary space was not homeomorphic to the elaston’s interior space. First elaston was a whole metric space and hence elaston was both closed and open (clopen). The boundary of this space is empty. A scalar field $\Lambda \sim a^{-2}$ (a - is a scale of non-uniformity of space) induced by metric is defined inside of elaston but on the complementary space this field is not determined since this space is non-metrizable. The energy density of elastic space is $c^4\Lambda/8\pi G$. The compressed space relaxes into Euclidean space that is filled up with particles in accordance with equation (1). The elastic energy that is contained in the elaston

$$E_{el} = (c^4/8\pi G) \oint_V \Lambda dV \qquad (10)$$

The physical space, when it is squeezed, possessed the elastic energy which gives birth to particles in the process of backward inflation. This is a modification of the Clifford's original idea with a distinction that in Clifford's case the physical entities themselves are manifestation of deformed space but in the present case the compressed space is a "pre-matter" and the particles are created at the expense of relaxation of the elastic stress.

The real elaston has to be spatially bounded. What is a boundary condition for cut-off of elaston? It is natural to consider as a boundary the distance where the density of elastic energy drops to the nuclear one: $\rho_{nucl} \sim 10^{14}$ g/sm$^3$.

To make flat space we need a phase transition on the boundary. The elastic energy is the free energy that produces particles and transforms into the rest mass energy. The entropy is a measure of the amount of energy in a physical system that can not be used to do the work. Then, by definition, an elaston has a minimal entropy and the entropy accumulated in evolution possessed by radiation and matter. The Euclidean space is a background vacuum state in contrast to the elaston that is an upper metastable state of 3-space. After this phase transition from elaston to vacuum state the space becomes a flat passive stage in the Newtonian sense. As a result the geometry acquires the most symmetrical space and physics gains particles and physical fields acting between them. After this the Ricci flow is unable to change the geometry of space. The flatness of our space is a consequence of this inability. The creation of particles which utilize the former elastic energy is an essential process to protect the energy conservation. Contrary to the cornerstone idea of GR the space becomes Euclidean in the presence of matter. The Ricci flow is a second order nonlinear parabolic differential evolution equation similar to the heat or diffusion equations

$$\frac{dg(t)}{dt} = -2Ricg(t) \qquad (11)$$

where g(t) is a metric, Ric g(t) is a tensor Ricci. The minus sign is because a plus one leads to backwards heat-type equation, which has no solution in general. The Ricci flow is the process which deforms given metric with variable Ricci curvature to even out of all the bumps and hollows in a manner formally analogous to diffusion or heat. The metric shrinks in positive Ricci curvature while it expands in the negative Ricci curvature direction.

Uniformization and geometrization are working together since for geometrization we need smooth space but smoothing implies geometry. There is no contradiction since geometrization here is regarded as Klein's conception of constant curvature that is characterized by its own group of isometries. General Riemannian geometry which includes spaces of variable curvature is much broader. The uniformization takes place on general Riemannian geometry and, only after the curvature of each decomposed homogeneous space becomes constant, space acquires geometry in Klein's meaning that is Thurston's geometrization. Does elaston have geometry? The answer is: *no* in the Klein's meaning and *yes* in the general Riemannian sense.

The evolution equation of the Ricci curvature has form

$$dR_{ij}/dt = \Delta R_{ij} + Q_{ij} \qquad (12)$$

where Δ is the Laplace-Beltrami operator, Q involves quadratic expressions in the curvature. This is a nonlinear heat-type equation. Analysis shows [20] that the Ricci flow preserves positive Ricci curvature in dimension 3 that is not a case for dimensions > 3. The Ricci flow does not preserve negative Ricci curvature in dimensions > 2. On the other hand the Ricci flow "geometrizes" 3-space of positive curvature. This analysis shows that the Ricci flow tends to favor positive curvature and even more, elastons ( Ric > 0) acquire a naturalness of existence only in 3-space. The dimension 3 of our physical space probably follows from this conclusion.

It is important that there are no geometries which interpolate continuously between the eight Thurston's geometric structures. The Ricci flow attempts to make the metric more homogeneous and isotropic, more symmetrical. The most symmetrical space is a flat Euclidean one. If the random Riemannian space has no boundary but is finite in size then Ricci flow, analogous to heat equation, can make homogeneous curvature but can not make it equal zero. That is geometry solely is not capable of making the more symmetric bounded flat space that is our World. It is precisely for this reason the elastic energy transmutes into the energy of the particles, to free the space from the energy that prevent it flattening.

When we use a term *evolution*, what do we mean? In quantum cosmology based on the Wheeler-DeWitt approach wavefunctional of the World Ψ satisfies the condition:

$$H\Psi = 0 \qquad (13)$$

that does not contain d/dt. That is an affirmation of the fact that we do not have the universal world time and therefore no any evolution of the World and its space. Here the emergence of geometry itself is caused by Ricci flow evolution equation. But when we say "evolution" we mean that space and time are the very different categories. The space is characterized by metric or Ricci tensor and it is a function of variable which we call "time" but, in general, in the differential geometry it may has nothing to do with any physical time. When we say "evolution" in respect to our real physical World we mean that parameter "t" in Ricci flow is *the physical time*. That means that space and time are entirely separate concepts. Is it correct that Ricci flow equation is a mathematical source that furnishes the clue to attribute a real physical time from the very outset? Parameter "t" may be determined as the physical time - "Ricci flow time" or RFT. So, within the elaston an evolution is determined by

$$\frac{dg(t)}{dt} = -2Ricg(t) \qquad (14)$$

Outside of elaston the evolution of the Euclidian volume of the World is determined by

$$\Pi\, \frac{dm}{dt} = \frac{dV}{dt}. \qquad (15)$$

RFT is running not only inside of the elaston but also outside since the boundary serves as a source of external space filled with particles. Therefore boundary plays a fundamental role. On the boundary of elaston the value of "t" must be sewed together. As a result "t" being an evolutionary parameter becomes the universal world time or what Newton called "absolute time" in contrast to the "referential" time of SR that depends of the speed of transmission of interaction between particles. Since some restructuring of space occurred before formation of the metrical space it is permissible to suppose existing of some form of time before RFT. However it is not possible now to comprehend what kind of physical reality it represented and how it had been connected to RFT.

There is sufficient difference of Ricci flow from the linear equations of diffusion or heat. It is precisely nonlinearity that causes singularities, that is typical for nonlinear equations. Here are two processes possible: a) smooth evolution of given space with no singularities, b) singularities forming with possible decomposition. If we have some general deformed 3-ball then it will be not easy to predict where the singularities will appear.

Fortunately, some important properties of Ricci flow depend on the qualitative considerations. Ricci flow converges, up to re-scaling, to metric of constant curvature, however "soliton" solutions to the flow give examples where the flow does not uniformize the metric, but only changes it by diffeomorphism as the time goes on. Soliton solutions are very important since they usually appear as blow up limits of flow. Some neck pinch singularities (with positive Euler characteristic) produce Ricci solitons. Ricci solitons arise as the limits of parabolic dilations about singularities. A Ricci soliton is generated by an initial metric g and a vector field V. A Riemannian metric is a gradient Ricci soliton if V = grad f with respect to g and its Hessian is equal to the Ricci tensor of g

$$\nabla^2 f = \mathrm{Ric}(g) \qquad (16)$$

Then the function f is called the Ricci potential. One important example of a gradient shrinking soliton is the Gaussian soliton. The known examples of Ricci solitons are the rotationally symmetric. It is showed in [21] that for 3-space there exist a complete, rotationally symmetric gradient soliton which metric cannot be written down explicitly. The problem of global existence and uniqueness of rotationally symmetric metric g with prescribed rotationally symmetric Ricci tensor

$$\mathrm{Ric}(g) = T \qquad (17)$$

on balls is examined in [22]. The conditions where found when this problem is solvable.

Sun-Chin Chu show [23] that spherically symmetric metric on 3-space is being a gradient Ricci soliton which positive curvature must be open like paraboloid. Scalar curvature has its maximum at the origin and falls off like 1/s.

In general, there are directions of positive and negative Ric(g) and metric may contract or expand. For dumbbell metric, for example, the neck may shrink since a positive curvature in $S^2$ direction will dominate [24]. After the neck is pinched off a topological decomposition into pieces happens. In [25] "corseting sphere" $S^3$ is numerically examined. A critical value of the corseting parameter was found. For small amount of corseting Ricci flow converges to the round sphere metric, while for a large amount of corseting a neck pinch singularity occurs. A pinching, probably, will happen also for the entity with more then one neck – multi-decomposition.

Here are two important corollaries for our model:

I. Ricci solitons, which are forming by evolution of Ricci flow, may be used as a mathematical description of elastons.

II. The elaston under supercritical regime may to experience neck pinch formation and decomposition of elaston may occur.

In accordance with our model a phase transition (curved space → flat space) is occur when moving boundary is formed. To use a Ricci flow technique we have to perform a transformation a(s, t) → R(s, t), Λ → R. Then we can formulate a boundary problem for a Ricci flow. The evolution equation for a scalar curvature

$$\frac{dR}{dt} = \Delta R + 2\left|Ric(g)\right|^2 \qquad (18)$$

is responsible for the elastic energy transfer in the elaston. Outside of elaston we have an equation (1) for the flows of new born mass with the rest mass energy and the volume of flat Euclidean space. A phase boundary can move with time and so this boundary problem is analogous to a Stefan problem for parabolic equation for free boundary motion in the flat space. The curvature outside R = 0 (flat space) and condition of phase transition on the boundary is

$$R\,\Big|_S \approx (8\pi G/c^2)\rho_{nucl}\,. \qquad (19)$$

Let us go back to physical space. We live in a finite World with a simply connected topology. Our World globally has composite geometry. It is a 3-ball that has Euclidean metric with immersed elastons that have non-Euclidean “compressed” metric. Euclidean space has a final geometry that is not liable more to the action of Ricci flow. An increase of Euclidean volume is the reason of expanding of our World. The elastons are mills for generation of new entities and Euclidean space is a calm zone for placid evolution of matter. This space differs from Newtonian global space which is too immovable and from Cartesian space that is too mobile. We make difference between space (stage) and particles (players) though in our comprehension a real Euclidean space exists (as a whole) only together with particles. This is not a relationism but **coexistence**. Here the stage and players come into being together: Euclidean space and particles cannot exist independently. Particles are the by-product of evolution of space. An evolution of the elastic space which was possessed the energy results in creation of final energetically passive but highly symmetrical flat space and the particles that reserve for themselves the former elastic energy.

We can look at the evolution of space from standpoint of the entropy. The evolution is bounded up with a change of the entropy. But change of the entropy is connected with a presence of coarse-graining structures. The fine-graining structures or uniform media conserve the entropy and so, not support evolution. It is precisely this fact that explains why our World had started in elastonic form. The elastons have high level of organization, minimal entropy S, minimal volume V and maximum reserve of free energy F, which is converting into the energy of the particles. At that, the new born Euclidean space itself has zero free energy. The Euclidean space is a ground state of the space condition. From the energy point of view the evolution of the space looks like following: free energy of space is diminishing $dG/dt < 0$, volume are increasing, $dV/dt > 0$. This means that evolution of our World is feeding by the free energy of elastons and is still elapsing at this epoch, as against to the Big Bang model where some instantaneous process happened long time ago and the results of which we are observing till now. New diverse structures emerge owing to the elastic energy of elastons. Our World is an open system that creates its order at the expense of the free energy leaking from elastons. This can happen only owing to the fact that the whole system of Euclidean section of space and elastons are far away from thermodynamic equilibrium.

The coarse-graining requires the presence of the boundaries. Just existing of the boundary of space evokes the sacred horror among the present-day cosmologists. They are trying somehow to twist the space so that any boundaries vanish. Of course, there are in mathematics some beautiful objects, for instance hypersphere or 3-sphere, without boundary but it is unlikely that something similar exists in our physical world. Sometimes it is said that Euclidean flat space can exist in two ways: infinite unbounded space or finite unbounded one in the form of flat 3-torus, but there is no any observational support for these models. Often the Cosmological principle or the Copernican principle is proposed to negate the boundary existence. But the Cosmological principle is not a law of nature; most probable it is a matter of belief. The Copernican World has the center – the Sun; so it is not compatible with homogeneous and isotropic world. An existence of the boundless world appears so inconceivable that possibility of existence of boundaries do not have to be encountered with hostility.

There two sorts of boundaries are available in our model and both are endowed with greatly important physical functions. We mentioned already the first kind of boundaries that separate elastons from Euclidean space. The second kind of boundaries is a border between Euclidean and non-metrizable space. In [11] we showed that this boundary has the properties of a black body with current apparent temperature ~ $3\,^{0}$ K. Let us recall the general considerations. Any photon incident upon the boundary may experience three different outcomes: reflection, transmission and absorption. A reflection is ruled out since there is no reflector. A transmission is not permitted owing to the fact that the space from the opposite side of the boundary is non-metric. The only possibility is an absorption being independent of frequency of the incident photon. This is exactly the property of the black body. The absorbed energy heats boundary to some effective temperature T that determines the temperature of the CMBR. So, the boundary of our World is the source of the CMBR. An observational equality of the CMBR energy density and the star radiation energy , otherwise mysterious, supports this model. Burbidge noted: "The fact that the energy density in the CMB is in good agreement with the amount of energy released in building the observed helium abundance through hydrogen burning. In 1926 Eddington made an order of magnitude estimate of the energy density of starlight and found a value of 7.67 x $10^{-13}$ erg cm$^{-3}$. Some later estimations give for the energy density of diffuse starlight in the universe is only of order $10^{-14}$ erg cm$^{-3}$. Let us take the average value $\rho \sim 10^{-13}$ erg cm$^{-3}$. This boundary is a skin layer for the incident electromagnetic radiation. It works like radiation thermalizer. The incident energy flow is $\rho \cdot c$ (erg cm$^{-2}$ sec$^{-1}$). If boundary under this flow of the energy acquires the temperature T then the blackbody radiation in accordance with Stefan - Boltzmann law is $j = \sigma T^4$. The balance of the energy is

$$\sigma T^4 = \rho \cdot c$$

and

$$T = \sqrt[4]{\frac{\rho c}{\sigma}} = \sqrt[4]{\frac{10^{-13} \cdot 3 \cdot 10^{10}}{5.7 \cdot 10^{-5}}} = 2.7 \text{ K}$$

So, the boundary acquires under the star light flow the temperature equal to the CMB temperature. It is hardly accidental coincidence.

This boundary is a skin layer for the incident electromagnetic radiation. It works like radiation thermalizer. Mathematically this boundary has properties both closed and open manifold. One section of space can be flat and others are curved. A metric space is trivially a gauge space and hence there exist some conservation laws of physical entities.

In our model the space on the outside of the elastons is always immutable Euclidean but not immovable like Newtonian space. The mutability of the space takes place only inside and on the boundary of elastons. The immutability of the final Euclidean space is a consequence of the property of the Ricci flow that has no effect on the flat space. The new born space that is determined by dV/dt locally is the absolute space in the Newtonian meaning, with the exception of “immovability”. That is the Newtonian space serves as the sections of the Cartesian “hydrodynamic” space.

**6 Physical space and micro world.**

The material world is made of protons, neutrons and electrons. The electrons contribute less than ~ 0.1 % of the mass. To understand an origin of the mass of surrounding world is actually to comprehend the origin of the mass of nucleon. In the quark-gluon model gluons are massless and the rest mass of quarks contributes ~ 1-2 % of the baryon mass. So almost the entire mass can be attributed to the quantum chromodynamics binding energy of gluons. How does this picture correspond to our idea of the reality of physical space?

We have in the elaston a compressed medium endowed with free elastic energy. Let us explore an analogy of decompactification of compressed space as a nozzle flow with the processes in the flowing liquid. When pressure in the flow decreases the nucleation occurs. There are two different homogeneous nucleation processes with the bubbles or drops creation, which requires superheating or supercooling. During nucleation a new phase is created, that requires the formation of an interface at the boundary between phases. We can introduce a concept of a surface tension σ and the difference $\Delta p = p_I - p$ of the interior pressure $p_I$ and exterior pressure p. Then Δp between two phases in equilibrium will relate to the radius of the nucleus r, by Laplace equation

$$\Delta p = 2\sigma/r \tag{20}$$

At the expense of elaston's free energy G a drop is created

$$\Delta G = 4\pi r^3\Delta p/3 + 4\pi r^2\sigma \tag{21}$$

where the first term is a volume energy and the second is a energy of surface tension of the new interface. The free energy for creation of drop in equilibrium is

$$\Delta G_{cr} = 16\pi\sigma^3/3(\Delta p)^2 \tag{22}$$

The supercooled media stops nucleation when free energy of decompactifing space is exhausted. A final Euclidean space has zero free energy and frozen nuclei that are the baryons. We can see from (22) that three parameters: $\Delta G_{cr}$, σ, Δp determine properties of the created particles. The value of $\Delta G_{cr}$ may determine the rest energy of emergent baryons. Then, suppose $r \sim 10^{-13}$ cm and $\Delta G_{cr} \sim 1.5*10^{-3}$ erg we find from (20) and (21), the order of magnitude $\sigma \sim 10^{23}$ erg/cm$^2$ and $\Delta p \sim 10^{36}$ erg/cm$^3$ $\sim 10^{30}$ atm.

To compare this model to quark-gluon model we can identify Δp with kinetic energy of the quarks and interface boundary between two phases with gluon field. So, the bag where quarks and gluons can exist is simply a boundary of drop of a new phase. *The confinement* of quarks has simple explanation as existence of a new phase of physical space living only in the interior of the drop. There is no internal energy to create new area of boundary. The asymptotic freedom follows from free non-viscous motion of the new phase limited only by the boundary that is a gluon-like force. The energy of the surface tension gives the overwhelming part of the mass of nucleon. In some sense there is not a space filled with the background Higgs field but physical space itself is an analog of this field. The inertia is a result of interaction of drops with the physical space. If drop moves uniformly trough perfect fluid (that is physical space) it does not experience a drag force (d'Alambert's paradox). If, however, we will try to accelerate a drop relative to physical space there "joined" mass emerges as a result of interaction with milieu. Now we can consolidate this new idea with conclusions we made in the part 4.

*The presence of the interface boundary in the physical space, possessing the energy of the surface tension $E_0$, caused two properties of the particle:*

*it has the equivalent mass $m = E_0/c^2$;*

*it has the inertia as a consequence of interaction with surrounding physical space that is a perfect fluid.*

The materiality is a property of the particles. They include the physical phase of space which does not exist in the surrounding milieu. There is a boundary dividing two phases , that possess the energy analogous to the energy of the surface tension. Just this energy is the rest energy of the particle $E_0$ and the equivalent to this energy mass is the mass of the particle $m = E_0/c^2$. Thereby the materiality, in our opinion, is an availability of the border within physical space that separates different spatial phases. The physical space that do not include particles is a real entity but this entity is not material. The particles are not the formations in the nothingness but they are non-uniformities of the physical space itself.

In the "common" interpretation the real entities are particles and physical fields but space is a void not entity. In our treatment particles and physical space are real entities. The physical fields are the forms of existence of the space trough which moves the energy in the form of quasi particles - bosons. So, both particles and fields the products of physical space. The physical space, as we mentioned, is the real entity so it could change the forms of its existence. The result of this process is the time. The time and evolution of space are inseparable. Evolution of space begets the course of time. The time is flowing in the direction assigned by the evolution of space which determines the arrow of time. But this evolution doesn't depend of somebody's will. Out of this evolution the time doesn't exist. The instruments which govern the time are variable properties of space and the energy that begets by evolution. For example, our world was born in the form of spatial domain with high free energy. This initial condition defined the existing of arrow of time. The emergence of particles and different types of energy is a byproduct of evolution. The space exists always; the particles and physical fields are transients.

The main constituent of mass - baryons, emerge on the stage of the sharp expansion of the "compressed" space on the borders of the elastons. This emergence is analogous to the formation of the liquid drops in the supercooled expanding gas. The elastic energy of "compressed" space is transformed into the energy of the surface tension of the boundaries of newly born particles.

One can speculate about nature of the other particles in this model. Most of the particles generating in the high energy collisions may be considered artificial products of superheated space medium when unstable nucleated bubbles are created. So, the proton and neutron are the natural outcome of phase transition in the supercooled flow of space from elaston. The antiparticles may also associate with opposite superheating that exist in artificially created condition like high energy collisions of existing particles. This process does not exist in the decompactification process that may be the explanation of the particle-antiparticle asymmetry. The drops emerges as a result of supercooling but bubbles as a result of superheating. We can not create a new baryon charge because our physical space is not a compressed space of elaston but a relaxed flat space which lost already its elastic energy. However by concentration of the energy in the very small region we can artificially "reallocate" physical space to create regions with "supercooled" and " superheated" phases. Accordingly we get "drop" - baryon and "bubble" -antibaryon. Both new born particles have analogous borders, what defines an equality their rest energies and respectively equality of masses. On their meeting they annihilate each other restoring initial physical space and the released energy is taken away by bosons. So, this is completely reversible process. As a result there is no creation of a new baryon charge. In some sense this picture remind the idea of "Dirac sea" where live particles and antiparticles - "holes". Thus there are no two drops occupying the same position, they are fermions. Bosons in this model are excitations of the physical space which transfer energy so they are quasi particles. Baryon-antibaryon asymmetry follows from existence of elastons which give birth to baryons on the border with the expansion and "supercooling" of space. The existence of "anti-elastons" with "superheating" is not foreseen. So, baryonic asymmetry is a consequence of evolution of space.

Interaction between physical space and movable particles depends of their relative velocity because physical space as an entity may have its own motion. Some regions of space can move relative to others. In our model the uniform relative motion of space and particles doesn't cause its mechanical interaction. Physical space behaves as an ideal fluid in this case. However non-uniform relative motion do cause interaction just as in the case of d'Alembert's sphere that has no relation to the viscosity of the fluid but is caused by asymmetry of the flow. Precisely this interaction is the source of inertia. To change relative velocity we need to apply external coercion to the particle or to the space. Thereby the source of inertia is an interaction of non-uniformly moving particle and physical space but not the interaction with remote matter as in the Mach's explanation.

As "the ether", the term "vacuum" has long and controversial history so its usage is too uncertain and multifunctional. A usage of both terms in this meaning engenders meaningless discussions.

Multi-nucleon systems may develop into nuclei by means of cohesion that is surface tension serving as the nuclear force or analog to the van der Waals interaction. The "wetting" cause clustering or agglomeration of nucleons. It is clear that a potential of surface force is short-acting that is consistent with property of nuclear force. Also, surface tension is charge independent. This picture is consistent with a liquid-drop model of nuclei. If internal stability of nucleons is stronger than cohesion force between them then we can explain the properties of Woods-Saxon potential for nuclear force

* it is attractive as a cohesion force.
* for large number of nucleons it is near flat in the center due to isotropy of the cohesion.
* nucleons near the surface of the nucleus experience a large force towards the center.
* it fast approaches zero as distance goes to infinity as cohesion is a short-acting.

The gravitational force is always attractive that is a property of cohesion. Naively we can suggest that gravity is "a weak tail" or residual nuclear force of the same interaction that leads to the nuclei formation. Gravity may falls off at large distances between galaxies faster than the inverse square law, that is in agreement with observations [26].

## 7 Concluding remarks

In this paper we briefly presented a new paradigm - a reality of the evolving physical space contrary to the standard view of the space as a purely relational nonexistence - void. In the next paper II we will present some qualitative and quantitative corollaries with comparisons to mostly astrophysical observations. The present approach, on the face of it, looks promising, though this is not a theory yet but rather an outline for the future development. Even if it has serious shortcomings it can stimulate an alternative developments in the physics and cosmology. We can only ask to read this text without bias and to continue to build upon it.